\documentclass[twocolumn,prl,showpacs,preprintnumbers,amsmath,amssymb]{revtex4}
\newcommand{\be}{\begin{equation}}
\newcommand{\ee}{\end{equation}}
\newcommand{\ba}{\begin{eqnarray}}
\newcommand{\ea}{\end{eqnarray}}

\usepackage{graphicx}
\usepackage{dcolumn}
\usepackage{bm}
\usepackage{color}

\begin{document}



\title{Effects of compression on the vibrational modes of marginally jammed solids}
\author{Matthieu Wyart}

\affiliation{Service de Physique de l'Etat Condens\'e (CNRS URA 2464), DSM/DRECAM, CEA Saclay,
91191 Gif sur Yvette, France}
\email{matthieu.wyart@m4x.org}
\author{ Leonardo E. Silbert, Sidney R. Nagel, Thomas A.  Witten}
\affiliation{ The James Frank Institute, The University of Chicago, Chicago, Illinois 60637}
\date{\today}

\begin{abstract}

  Glasses have a large excess of low-frequency vibrational modes in
  comparison with most crystalline solids.  We show that such a
  feature is a necessary consequence of the weak connectivity of the
  solid, and that the frequency of modes in excess is very sensitive
  to the pressure. We analyze in particular two systems whose density
  $D(\omega)$ of vibrational modes of angular frequency $\omega$
  display scaling behaviors with the packing fraction: (i) simulations
  of jammed packings of particles interacting through finite-range,
  purely repulsive potentials, comprised of weakly compressed spheres
  at zero temperature and (ii) a system with the same network of
  contacts, but where the force between any particles in contact (and
  therefore the total pressure) is set to zero. We account in the two
  cases for the observed a) convergence of $D(\omega)$ toward a
  non-zero constant as $\omega\rightarrow 0$, b) appearance of a
  low-frequency cutoff $\omega^*$, and c) power-law increase of
  $\omega^*$ with compression. Differences between these two systems
  occur at lower frequency. The density of states of the modified
  system displays an abrupt plateau that appears at $\omega^*$, below
  which we expect the system to behave as a normal, continuous,
  elastic body. In the unmodified system, the pressure lowers the
  frequency of the modes in excess. The requirement of stability
  despite the destabilizing effect of pressure yields a lower bound on
  the number of extra contact per particle $\delta z$: $\delta z \geq
  p^\frac{1}{2}$, which generalizes the Maxwell criterion for rigidity
  when pressure is present. This scaling behavior is observed in the
  simulations. We finally discuss how the cooling procedure can affect
  the microscopic structure and the density of normal modes.

\end{abstract}
\maketitle 

\section{I Introduction}

The responses to thermal or mechanical excitations and the transport
properties of a solid depend on its low-frequency vibrations. In a
continuous elastic body, translational invariance requires that the
vibrational modes, or normal modes, are acoustic modes. This leads,
notably, to the Debye behavior of the density of vibrational modes
$D(\omega)$: in three dimensions, $D(\omega)\sim \omega^2$
\cite{Ashcroft}. In amorphous solids structural disorder is present
and this description breaks down at intermediate frequencies and
length scales. In glasses, there is a large excess of
low-frequency vibrations observed in neutron scattering experiments.
This excess corresponds to a broad maximum in $D(\omega)/\omega^2$,
called the``boson peak", which appears at frequencies on the order of
a terahertz \cite{malinovsky}, that is typically between one hundredth
and one tenth of the Debye frequency.  Above this frequency transport
properties \cite{W.Phillips} are strongly affected, as in silica where
the thermal conductivity is several orders of magnitude smaller than in
a crystal of identical composition \cite{bergman}. Furthermore,
amorphous solids also exhibit force inhomogeneities that are common to
granular materials \cite{ohern} and glasses \cite{tanguy}.  Thus,
the resulting force propagation due to a localised perturbation
differs considerably from the predictions of continuum elasticity
theory at short length scales. Recent simulations reported that this
difference appears below a characteristic length scale which is
directly related to the frequency of the boson peak \cite{tanguy}.
Thus the excess-modes play an important role in several anomalous
properties of amorphous solids.  It was recently proposed that they
also govern the physics at the liquid-glass transition \cite{parisi}.

Despite their ubiquity, there is no accepted explanation of the underlying
cause of these excess vibrational modes in glasses.  A dramatic illustration of this excess 
was found in recent computer simulations of soft-spheres
with repulsive, finite range potentials near the jamming transition \cite{J}. At this transition
an amorphous solid loses both its bulk and shear moduli and
becomes a liquid \cite{liu}. In a recent letter \cite{me} we showed
how to calculate the density of states for weakly-connected amorphous
solids, such as those near jamming, and showed that an
excess density of vibrational states is a necessary feature of such
systems. In this paper we use the method of \cite{me} to predict
further consequences of compression. In particular, we explain why the
coordination $z$ must increase in a non-analytic way with applied
pressure, as observed in the simulations \cite{durian,ohern,J}.

\begin{figure}

\centering
\includegraphics[angle=0,width=7cm]{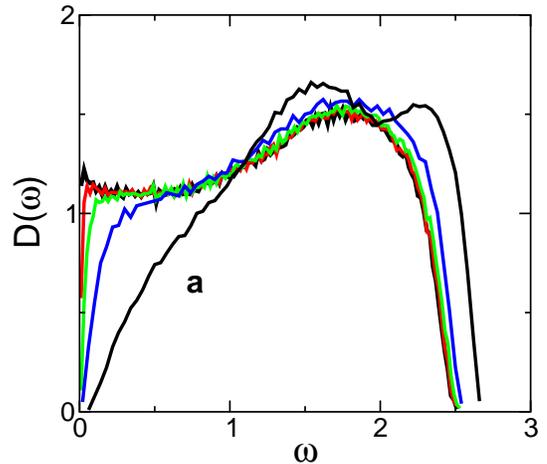}
\caption {The density of vibrational states, $D(\omega)$, vs angular frequency, $\omega$, for the simulation of Ref.~\cite{J}.  1024 spheres interacting with repulsive harmonic potentials were compressed in a periodic cubic box to packing fraction  $\phi$, slightly above the jamming threshold $\phi_c$.  Then the energy for arbitrary small displacements was calculated and the dynamical matrix inferred. The curve labeled $a$ is at a relative packing fraction $\phi - \phi_c = 0.1$.  Proceeding to the left the curves have relative volume fractions $10^{-2}$,  $10^{-3}$,  $10^{-4}$,  $10^{-8}$, respectively.  }
\label{f1}
\end{figure}

Because there are no attractive forces and the temperature is zero in the soft-sphere simulations of O'Hern et al.
\cite{ohern,J,nagel}, the pressure $p=0$ at the packing fraction at the jamming
transition, $\phi_c$.  In 3-dimensions for mono-disperse spheres, $\phi_c
\approx 0.64$.  The average number of contacting neighbors per
particle, $z$, and the elastic moduli scale as functions of $p$.  These simulations also reveal unexpected features in
the density of vibrational modes, $D(\omega)$: (a) As shown in
Fig.~(\ref{f1}), when the system is at the limit of marginal
stability, as $p\rightarrow 0$, $D(\omega)$ has a plateau extending
down to zero frequency with no sign of the standard $\omega^2$ density
of states normally expected for a three-dimensional solid.  (b) The
plateau is progressively eroded at frequencies below a characteristic
frequency $\omega^*$, that increases with the pressure $p$ (see
Fig.~\ref{sans}). (c) The value of $D(\omega)$ of the plateau is
unaffected by this compression. (d) At frequencies much lower than
$\omega^*$, $D(\omega)$ still increases much faster with $\omega$ than
the quadratic Debye dependence. Similar behavior was also recently seen in
models of tetrahedral covalent glasses \cite{dove}.  An earlier
simulation of a Lennard-Jones glass had indicated that $D(\omega)$
increases at low frequencies when the coordination number, $z$, is
lowered \cite{Grest}.  In the following we aim to relate the density
of vibrational modes of weakly-connected amorphous solids such as
an assembly of finite-range, repulsive particles to their microscopic structure. We
start by briefly reviewing the results of \cite{me} where the scaling
properties of the density of states were derived.  Then we use these
results to compute the effect of applied stress. We show that it leads
to a non-trivial constraint on the geometry of the contact network near
the jamming threshold. The scaling behavior of the soft spheres near
jamming furnish a stringent test to our predictions.

At the center of our argument lies the concept of {\it soft modes}, or
{\it floppy modes}. These are collective modes that conserve the
distance, at first order, between any particles in contact.  They have
been discussed in relation to various weakly-connected networks such
as covalent glasses \cite{phillips, thorpe}, Alexander's models of
soft solids \cite{shlomon}, models of static forces in granular packs
\cite{Tom1,moukarzel} and rigidity percolation models, see e.g.
\cite{yeye}.  As we shall discuss below, they are present when a
system has too low a coordination number.  As a consequence, as Maxwell showed
\cite{max}, a system with a low average coordination number $z$ has
some soft modes and is therefore not rigid. There is a threshold value
$z_c$ where a system can become stable, such a state is called {\it
  isostatic}.  As we shall discuss this is the case at the jamming
transition, if rattlers (particles with no contacts) are excluded.
At this point there are no zero-frequency modes except for the trivial translation
modes of the system as a whole.  However, if any contact were to be
removed, the frequency of one mode would go to zero, that is, one soft
mode would appear.  As we argued in \cite{me}, this idea can be used to show that
isostatic states have a constant density of states in any dimensions.  When $z > z_c$, the system still behaves
as an isostatic medium at a short enough length scale, which leads to the
persistence of a plateau in the density of states at intermediate frequency.

The second concept we use is at the heart of
Alexander's discussion of soft solids \cite{shlomon}.  In continuum elasticity the
expansion of the energy for small displacements contains a term
proportional to the applied stress. It is responsible for the
vibrations of strings and drumheads and also for inelastic
instabilities such as the buckling of thin rods. Alexander pointed out
that this term also has a strong effect at a microscopic level in
weakly-connected solids. For example, it confers rigidity to gels,
even though these do not satisfy the Maxwell criterion for rigidity.
We will show that while this term does not greatly affect the acoustic
modes, it nevertheless strongly affects the soft modes. In a repulsive system of
spherical particles it lowers their frequency. We argue that this can
dramatically change the density of states at low frequency, as
confirmed by a comparison of simulations where the force in any
contact is present, or set to zero. We show that these considerations
also lead to a relation between the excess connectivity $\delta
z\equiv z-z_c$ and the pressure, $p$.

The paper is organized as follows. In the next section we write the
expansion of the elastic energy that we use to derive the soft-modes
equation. We then define and discuss the isostatic case, and the
nature of the soft modes that appear when contacts are removed in
such a system. In the third part we compute the density of states
when the effect of the applied stress on the vibrations is neglected.
This approximation corresponds to a real physical system: a network of
relaxed springs. We use the soft modes in a variational argument to
show that an isostatic state has a constant density of vibrational
modes. We extend this argument to include the case where where the coordination number increases above
$z_c$, as is the case of the soft-sphere system
under isotropic compression. We show that such a system behaves as an
isostatic one for length scales smaller than $l^*\sim \delta z^{-1}$.
This leads to a plateau in the density of states for frequency higher
than $\omega^*\sim \delta z$. At lower frequencies we expect the
system to behave as a continuous medium with a Debye regime, which is
consistent with our simulations. In the fourth part we study the effect
of an applied pressure on $D(\omega)$. We show that although it does
not affect the acoustic modes, it lowers the frequency of the soft
modes. We give a simple scaling argument to evaluate this effect, and
discuss its implication for the density of states. Incidentally this
also furnishes an inequality between $\delta z$ and the pressure,
which is also verified by the simulations, and which generalizes the
Maxwell criterion for rigidity. Finally we discuss the influence of
the cooling rate and the temperature history on the spatial structure
and the density of states of the system, and conclude our work.

\section{II Soft Modes and Isostaticity}

\subsection{ Energy expansion}

Following \cite{J} we consider $N$ soft spheres packed into a
spatially periodic cubic cell of side $L$ at volume fraction $\phi$.
To describe the vibrations of a system of particles, we expand the
energy around equilibrium. For a central interaction $V(r)$ one
obtains:

\be
\label{0}
\delta E=  \sum_{ij} V'(r_{ij}^{eq}) dr_{ij}+ \frac{1}{2} V''(r_{ij}^{eq}) dr_{ij}^2+ O(dr_{ij}^3)
\ee
where the sum is over all pairs of particles, $r_{ij}^{eq}$ is the equilibrium distance between particles $i$ and $j$. In order to get an expansion in terms of the particle displacements from equilibrium $\delta
\vec{R_1}... \delta
\vec{R_N}$  we use:

\be
\label{00}
dr_{ij}= (\delta\vec{R_j}-\delta\vec{R_i}).\vec n_{ij}+ \frac{[(\delta\vec{R_j}-\delta\vec{R_i})^{\bot}]^2}{2 r_{ij}^{eq}}+
O(\delta\vec{R}^3)
\ee
where $\vec n_{ij}$ is the unit vector along the direction $ij$, and $ (\delta\vec{R_j}-\delta\vec{R_i})^{\bot}$ indicates the projection of $\delta\vec{R_j}-\delta\vec{R_i}$ on the plane orthogonal to $\vec n_{ij}$.  When Eq.~(\ref{00}) is used in Eq.~(\ref{0}), the linear term in the displacement field disappears (the system is at equilibrium) and we obtain:

\ba
\label{000}
\delta E= \sum_{ij} V'(r_{ij}^{eq}) \frac{[(\delta\vec{R_j}-\delta\vec{R_i})^{\bot}]^2}{2 r_{ij}^{eq}} \\
+ \frac{1}{2} V''(r_{ij}^{eq}) [(\delta\vec{R_j}-\delta\vec{R_i}).\vec n_{ij}]^2+O(\delta\vec{R}^3)
\ea
 In what follows we  consider repulsive, finite-range ``soft spheres''.  For inter-particle distance $r<\sigma$, the particles have non-zero mutual energy and are said to be in contact. They  interact with the following potential:

\be
\label{opo}
 V(r)=\frac{\epsilon}{\alpha} \left(1-\frac{r}{\sigma}\right)^{\alpha}
 \ee
  where $\sigma$ is the particle diameter  and $\epsilon$ a characteristic energy.  For $r>\sigma$  the potential vanishes and particles do not interact.  Henceforth we express all distances in units of $\sigma$, all energies in units of $\epsilon$, and all masses in units of the particle mass, $m$.   In the following, we consider the harmonic case $\alpha=2$.  In the discussion section we argue that these results can be extended for example to the case of Hertzian contacts \cite{J} where $\alpha = 5/2$.  In the harmonic case we have:

\ba
\label{1}
\delta E=  [ \frac{1}{2}\sum_{\langle  ij \rangle} (r_{ij}^{eq}-1) \frac{[(\delta\vec{R_j}-\delta\vec{R_i})^{\bot}]^2}{2 r_{ij}^{eq}} ] \notag
\\
+  \frac{1}{2}\sum_{\langle  ij \rangle} [(\delta\vec{R_j}-\delta\vec{R_i}).\vec n_{ij}]^2+O(\delta \vec{R}^3)
\ea
where the sum is over all $N_c$ contacts $\langle ij\rangle$. It is convenient to express Eq.~(\ref{1}) in matrix form, by defining the set of displacements $\delta \vec R_1 ... \delta \vec R_N$ as a $dN$-component vector
$|\delta {\bf R}\rangle$.  Then Eq.~(\ref{1}) can be written in the form
$\delta E = 
\langle\delta {\bf R}| {\cal M}|\delta {\bf R}\rangle$.  The corresponding
matrix ${\cal M}$ is known as the dynamical matrix \cite{Ashcroft}.  The $dN$
eigenvectors of the dynamical matrix are the normal modes of the particle
system, and its eigenvalues are the squared angular frequencies of these modes.

The first term in Eq.~(\ref{1}) is proportional to the contact forces. In the rest of this paper we shall refer to this term as the {\it stress term} or {\it transverse term}.  Near the jamming transition $ r_{ij}^{eq} \rightarrow 1$ so that this term becomes arbitrarily small. We start by neglecting it, and we come back to its effects in the last section. This approximation corresponds to a real physical system where the soft spheres are replaced by point particles interacting with relaxed springs. We now have:

\ba
\label{2}
\delta E= \frac{1}{2} \sum_{\langle  ij \rangle}[(\delta\vec{R_j}-\delta\vec{R_i}).\vec n_{ij}]^2  
\ea
${\cal M}$  can be written as an $ N$ by $N$ matrix  whose elements are themselves tensors of rank $d$, the spatial dimension:
\be
{\cal M}_{ij}=-\delta_{\langle ij \rangle} \vec{n_{ij}}\otimes \vec{n_{ij}} + \delta_{i,j} \sum_{<l>}\vec{n_{il}} \otimes \vec{n_{il}}
\ee
 where $\delta_{\langle ij \rangle}=1$ when i and j are in contact and the sum is taken on all the contacts $l$ with $i$.

\subsection {Soft Modes}

If the system has too few contacts, $\cal M$  has a set of modes of vanishing restoring force and thus vanishing vibrational frequency.  These are the {\it soft modes} mentioned in the introduction. For these soft modes the energy $\delta E$ of Eq.~(\ref{2}) must vanish; therefore they must satisfy the $N_c$ constraint equations:
\be
\label{3}
(\delta\vec{R_i}-\delta\vec{R_j}).\vec n_{ij}=0 \ \hbox{ for all $N_c$ contacts}\ \langle ij \rangle
\ee
This linear equation defines the vector space of displacement fields that conserve the distances at first order between particles in contact. The particles can yield without restoring force if their displacements lie in this vector space.  Eq.~(\ref{3}) is purely geometrical and does not depend on the interaction potential. Each equation restricts the $dN$-dimensional space of $|\delta{\bf R}\rangle$ by one dimension.  In general, these dimensions are independent, so that the number of independent soft modes is $dN - N_c$. Of these, $d(d+1)/2$ modes are dictated by the translational and rotational invariance of the energy function $\delta E[\delta {\bf R}]$. Apart from these, there are $dN - N_c - d(d+1)/2$ independent internal soft modes.

\subsection{Isostaticity}

There are no internal soft modes in a rigid solid. This is true when a system of repulsive spheres jams, when the rattlers ({it i.e.}, particles without contacts) are removed. Therefore jammed states must satisfy  $N_c \geq dN-d(d+1)/2$, which is the Maxwell criterion for rigidity.  In fact at the jamming transition, if rattlers \footnote[1]{At the jamming transition the rigid system is a d-dimensional object with only a few holes and the rattlers are typically clusters of one or two particles. This is very different from the rigidity percolation models where bonds are randomly deposited on a lattice.  In that case, the percolating rigid cluster is a fractal object with dimension smaller than d.} are removed, one can show this inequality becomes an equality \cite{Tom1,moukarzel,roux}, as was verified in \cite{ohern}.  Such a system is called {\it isostatic}. The coordination number $z$ is then $z_c\equiv 2N_c/N \rightarrow 2d$. 

\begin{figure}
\centering
\includegraphics[width=7cm]{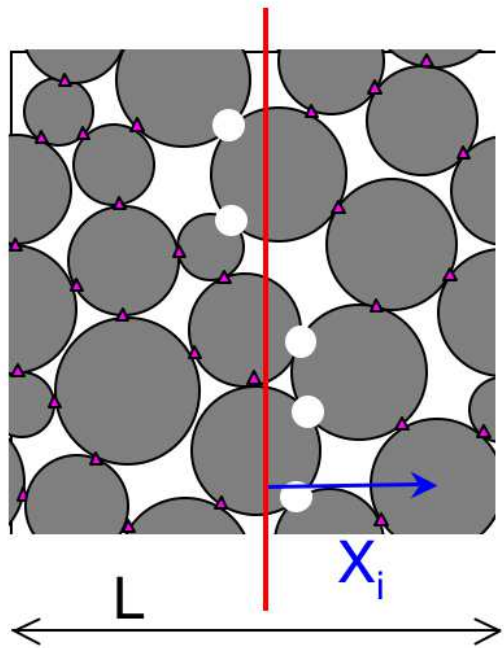}
\caption{Illustration of the boundary contact removal process described in the text.  Eighteen particles are confined in a square box of side $L$ periodically continued horizontally and vertically.  An isostatic packing requires 33 contacts in this two-dimensional system.  An arbitrarily drawn vertical line divides the system.  A contact is removed wherever the line separates the contact from the center of a particle.   Twenty-eight small triangles mark the intact contacts; removed contacts are shown by the five white circles.  }
\label{figs}
\end{figure}

\begin{figure}

\centering
\includegraphics[width=7cm]{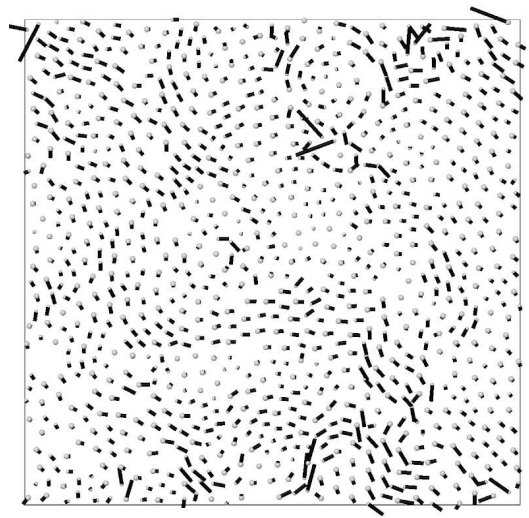}
\caption{ One soft mode in two dimensions for $N\approx 1000$ particles.  Each particle is represented by a dot. The relative displacement of the soft mode is represented by a line segment extending from the dot. The mode was created from a previously prepared isostatic configuration, periodic in both directions, following \cite{J}.  20 contacts along the vertical edges were then removed and the soft modes determined.  The mode pictured is an arbitrary linear combination of these
modes.}  
\label{softfig}
\end{figure}

An isostatic system is marginally stable: if $q$ contacts are cut, a
space of soft modes of dimension $q$ appears. For our coming argument
we need to discuss the extended character of these modes. In general
when only one contact $\langle ij \rangle$ is cut in an isostatic
system, the corresponding soft mode is not localized near $\langle ij
\rangle$. This arises from the non-locality of the isostatic condition
that gives rise to the soft modes; as was confirmed in the isostatic
simulations of Ref.~\cite{Tom1}, which observed that the amplitudes of
the soft modes were spread over a nonzero fraction of the particles.
When many contacts are severed, the extended character of the soft
modes that appear depends on the geometry of the region being cut. If
this region is compact many of the soft modes are localized.  For
example cutting all the contacts inside a sphere totally disconnects
each particle within the sphere. Most of the soft modes are then the
individual translations of these particles and are not extended
throughout the system.

In what follows we will be particularly interested in the case where
the region of the cut is a hyperplane as illustrated in
Fig.~(\ref{figs}). In this situation occasionally particles in the
vicinity of the hyperplane can be left with less than $d$ contacts, so
that trivial localized soft modes can also appear. However we expect
that there is a non-vanishing fraction $q'$ of the total soft modes
that are not localized near the hyperplane, but rather extend over the
entire system, like the mode shown in Fig.~(\ref{softfig}).  We shall
define extended modes more precisely in the next section.

\section{ III $D(\omega)$ in a system of relaxed springs}
\label{to}
\subsection{ Isostatic case}

\subsubsection {Variational procedure}

We aim to show first that the density of states of an isostatic system does not vanish at zero frequency. $D(\omega)$ is the total number of modes per unit volume per unit frequency range.  Therefore we have to show that there are at least on the order of $\omega L^d$ normal modes with frequencies smaller than $\omega$ for any small $\omega$ in a system of linear size $L$. As we justify later, if proven in a system of size $L$ for $\omega\sim \omega_L\sim 1/L$, this property can be extended to a larger range of $\omega$ independant of $L$. Therefore it is sufficient to show that they are of the order of $L^{d-1}$ normal modes with frequency of the order of $1/L$, instead of the order of one such mode in a continuous solid. 

To do so we use a variational argument: ${\cal M}$ is a positive symmetric matrix. Therefore if a normalized mode has an energy $\delta E$, we know that the lowest eigenmode has a frequency $\omega_0\leq \sqrt {\delta E}$. Such argument can be extended to a set of modes \footnote[2]{ If $m_\alpha$ is the $\alpha$'th lowest eigenvalue of ${\cal M}$ and if $e_\alpha$ is an orthonormal basis such that $\langle e_\alpha|{\cal M}| e_\alpha\rangle \equiv n_\alpha$ then the variational bound of A. Horn [Am. J. Math {\bf 76} 620 (1954)] shows that $\sum_1^q m_\alpha \leq \sum_1^q n_\alpha$.  Since $q n_q \geq \sum_1^q n_\alpha$, and since $\sum_1^q m_\alpha \geq \sum_{q/2}^q m_\alpha \geq (q/2) m_{q/2}$, we have $q m_p \geq (q/2) n_{q/2}$ as claimed.}: if there are $m$ {\it orthonormal} trial modes with energy $\delta E \leq \omega_t^2$, then there are at least $m/2$ eigenmodes with frequency smaller than $\sqrt 2\omega_t$. Therefore we are led to find of the order of $L^{d-1}$ trial orthonormal modes with energy of order $ 1/L^2$.

\subsubsection{Trial modes}

Our procedure for identifying the lowest frequency modes resembles that used for an ordinary solid.  An isolated block of solid has three soft modes that are simply translations along the three co-ordinate axes.  If the  block is enclosed in a rigid container, translation is no longer a soft mode.  However, one may recover the lowest-frequency, fundamental modes by making a smooth, sinusoidal distortion of the original soft modes.  We follow an analogous procedure to find the fundamental modes of our isostatic system.  First we identify the soft modes associated with the boundary constraints by removing these constraints.  Next we find a smooth, sinusoidal distortion of these modes that allows us to restore these constraints.
    
For concreteness we consider the three-dimensional cubical $N$-particle system $\cal S$ of Ref.~\cite{J} with periodic boundary conditions at the jamming threshold.  We label the axes of the cube by x, y, z.  $\cal S$ is isostatic, so that the removal of $n$  contacts allows exactly $n$ displacement modes with no restoring force. Consider for example the system $\cal S'$ built from $\cal S$ by removing the $q\sim L^2$ contacts crossing an arbitrary plane orthogonal to (ox); by convention at $x=0$, see Fig.~(\ref{figs}).  $\cal S'$, which has a free boundary condition instead of periodic ones along (ox), contains a space of soft modes of dimension $q$\footnote[3] { The balance of force can be satisfied in ${\cal S'}$ by imposing external forces on the free boundary. This adds a linear term in the energy expansion that does not affect the normal modes.}, instead of one such mode ---the translation of the whole system--- in a normal solid.  As stated above, we suppose that a subspace of dimension $q'\sim L^2$ of these soft modes contains only extended modes.  We define the {\it extension} of a mode relative to the cut hyperplane in terms of the amplitudes of the mode at distance $x$ from this hyperplane.  Specifically the extension $e$ of a normalized mode $|\bf \delta R \rangle$ is defined by $\sum_i\sin^2(\frac{x_i \pi}{L}) \langle i|{\bf \delta R} \rangle^2 =e$, where the notation $ \langle i|{\bf \delta R} \rangle$ indicates the displacement of the particle $i$ of the mode considered.  For example, a uniform mode with $ \langle i|{\bf \delta R} \rangle$  constant for all sites has $e=1/2$ independent of L. On the other hand, if $ \langle i|{\bf \delta R} \rangle=0$ except for a site $i$ adjacent to the cut hyperplane, the $x_i/L\sim L^{-1}$ and $e\sim L^{-2}$. We define the subspace of extended modes by setting a fixed threshold of extension $e_0$ of order 1 and thus including only soft modes $\beta$ for which $e_\beta>e_0$. As we discussed in the last section, we expect that a fixed fraction of the soft modes remain extended as the system becomes large. Thus if $q'$ is the dimension of the  extended modes vector space, we shall suppose that $q'/q$ remains finite as $L\rightarrow \infty$. The appendix presents our numerical evidence for this behavior.

We now use the vector space of dimension $q'\sim L^2$ of extended soft modes of $\cal{S'}$ to build $q'$ orthonormal trial modes of $\cal{S}$ of frequency of the order $1/L$.  Let us define $|\bf \delta R_\beta \rangle$ to be a normalized basis of this space, $1\leq \beta \leq q'$.  These modes are not soft in the jammed system $\cal S$ since they deform the previous $q$ contacts located near $x=0$.  Nevertheless a set of trial modes, $|\bf \delta R_\beta^* \rangle$, can still be formed by altering the soft modes so that they do not have an appreciable amplitude at the boundary where the contacts were severed.  We seek to alter the soft mode to minimize the distortion at the severed contacts while minimizing the distortion elsewhere. Accordingly, for each soft mode $\beta$ we define the corresponding trial-mode displacement $\langle i|{\bf \delta R}^* \rangle$ to be: 
\be
\label{rr}
\langle i|{\bf \delta R}_\beta^* \rangle \equiv C_\beta \sin(\frac{x_i \pi}{L}) \langle i|\bf \delta R_\beta \rangle
\ee
where the constants $C_\beta $  are introduced to normalize the modes.   $C_\beta $ depends of the spatial distribution of the mode $\beta$. If for example, a highly localized mode has $ \langle i|{\bf \delta R} \rangle=0$ except for a site $i$ adjacent to the cut plane, $C_\beta$ grows without bound as $L\rightarrow \infty$.  In the case of extended modes $C_\beta^{-2}\equiv \sum_{\langle ij \rangle} \sin^2(\frac{x_i \pi}{L})   \langle j|{\bf \delta R}_\beta \rangle^2= e_\beta > e_0$, and  therefore $C_\beta$ is bounded above by  $e_0^{-1/2}$.  The sine factor suppresses the problematic gaps and overlaps at the $q$ contacts near $x=0$ and $x=L$. The unit basis $|\bf \delta R_\beta \rangle$ can always be chosen such that the  $|\bf \delta R_\beta^* \rangle$ are orthogonal, simply because the modulation by a sine that relates the two sets is an invertible linear mapping in the subspace of extended modes.  Furthermore one readily verifies that the energy of each $|\bf \delta R_\beta^* \rangle$ is small, and that the sine modulation generates an energy of order $1/L^2$ as expected. Indeed we have from Eq.~(\ref{2}):

\be
\delta E =  C_\beta^2 \sum_{\langle ij \rangle} [(\sin(\frac{x_i \pi}{L}) \langle i|{\bf \delta R}_\beta \rangle- \sin(\frac{x_j \pi}{L}) \langle j|{\bf \delta R}_\beta \rangle) \cdot \vec n_{ij}]^2
\ee
Using  Eq.~(\ref{3}), and expanding the sine, one obtains:

\ba
\delta E \approx  C_\beta^2 \sum_{\langle ij \rangle} \cos^2(\frac{x_i \pi}{L}) \frac{\pi^2}{L^2} (\vec n_{ij} \cdot \vec e_x)^2 ( \langle j|{\bf \delta R}_\beta \rangle \cdot \vec n_{ij})^2 \\
\label{kk}
\leq  e_0^{-1}  (\pi/L)^2 \sum_{\langle ij \rangle}\langle j|{\bf \delta R}_\beta \rangle^2
\ea
where $\vec e_x$ is the unit vector along (ox), and where we used $|\cos| \leq 1$. The sum on the contacts can be written as a sum on all the particles since only one index is present in each term. Using the normalization of the mode $\beta$ and the fact that the coordination number of a sphere is bounded by a constant $z_{max}$ ($z_{max}=12$ for 3 dimensional spheres \footnote[12]{ In a polydisperse system $z_{max}$ could a priori be larger. Nevertheless Eq.~(\ref{kk}) is a sum on every contact where the displacement of only one of the two particles appears in each term of the sum. The corresponding particle can be chosen arbitrarily. It is convenient to choose the smallest particle of each contact. Thus when this sum on every contact is written as a sum on every particle to obtain Eq.~(\ref{kkk}), the constant $z_{max}$ still corresponds to the monodisperse case, as a particle cannot have more  contacts than that with particles larger than itself. }. ), one obtains:

\be
\label{kkk}
\delta E\leq   e_0^{-1}  (\pi/L)^2 z_{max} \equiv \omega_L^2
\ee
We have found on the order of $L^2$ trial orthonormal modes of frequency bounded by $\omega_L\sim 1/L$, and we can apply the variational argument mentioned above: the average density of states is bounded below by a constant below frequencies of the order $\omega_L$.  In what follows, the trial modes introduced in Eq.~(\ref{rr}), which are the soft modes modulated by a sine wave, shall be called ``anomalous modes''.  

To conclude, one may ask if this variational argument can be improved, for example by considering geometries of broken contacts different from the hyperplane surfaces we have considered so far.  When contacts are cut to create a vector space of extended soft modes, the soft modes must be modulated with a function that vanishes where the contacts are broken in order to obtain trial modes of low energy. On the one hand, cutting many contacts increases the number of trial modes. On the other hand, if too many contacts are broken, the modulating function must have many ``nodes'' where it vanishes. Consequently this function displays larger gradients and the energies of the trial modes increase. Cutting a surface (or many surfaces, as we shall discuss below) is the best compromise between these two opposit effects. Thus our argument gives a natural limit to the number of low-frequency states to be expected.

\subsubsection{ Extension to a wider range of frequencies}

We may extend this argument to show that the bound on the average density of states extends to higher frequencies.  If the cubic simulation box were now divided into $m^3$ sub-cubes of size $L/m$, each sub-cube must have a density of states equal to the same $ D(\omega)$ as was derived above, but extending to frequencies on order of $m\omega_L$.  These subsystem modes must be present in the full system as well, therefore the  bound on $D(\omega)$ extends to $[0,m \omega_L]$.  We thus prove that the same bound on the average density of states holds down to sizes of the order of a few particles, corresponding to frequencies independent of $L$.  We note that in $d$ dimensions this argument may be repeated to yield a total number of modes, $L^{d-1}$, below a frequency $\omega_L \approx 1/L$, thus yielding a limiting nonzero density of states in any dimension.

We note that the trial modes of energy $\delta E\sim l^{-1}$ that we introduce by cutting the full system into subsystems of size $l$ are, by construction, localized to a distance scale $l$. Nevertheless we expect that these trial modes will hybridize with the trial modes of other, neighboring, subsystems; the corresponding normal modes will therefore not to be localized to such short length scales.       

\begin{figure}
\centering
\includegraphics[angle=0,width=7cm]{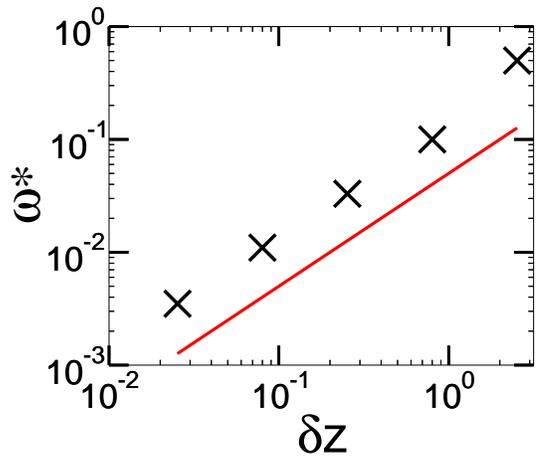}
\caption{Scaling of $\omega^*$ with the excess coordination number, $\delta z$ in the system with relaxed springs. The line has a slope one.}
\label{sans}
\end{figure}

\subsection{State with $\delta z>0$ }

When the system is compressed and moves away from the jamming transition, the simulations show that the extra-coordination number  $\delta z \equiv z-z_c$ increases. In the simulation, the compression also increases the force at all contacts. However, in this section we will ignore these forces and focus our attention only on the contact network created by the compression.  Any tension or compression in the contacts is removed. The effect on the energy is to remove the first bracketed term in Eq.~(\ref{1}) above.  We note that removing these forces, which must add to zero on each particle, does not disturb the equilibrium of the particles or create displacements.  In this section we ignore the question of how $\delta z$ depends on the degree of compression. We will return to this question in the next section.  

Compression causes $\Delta N_c=N \delta z/2 \sim L^d \delta z$ extra constraints to appear in Eq.~(\ref{2}).  Cutting the boundaries of the system, as we did above, relaxes $q\sim L^{d-1}$ constraints.  For a large system, $L^d\delta z> L^{d-1}$ and thus $q<\Delta N_c$.  Thus Eq.~(\ref{2}) is still over-constrained and there will be no soft modes in the system.  
However, as the systems become smaller, the excess number of constgraints, $\Delta N_c$, diminishes; for $L$ smaller than some $l^*\sim \delta z^{-1}$, $q$ becomes larger than $\Delta N_c$the system is again under-constrained as was already noticed in \cite{Tom1}.  This allows one to build low-frequency modes in subsystems smaller than $l^*$.  These modes appear above a cut-off frequency
$\omega^*\sim l^*{}^{-1}$; they are the ``anomalous modes" that contribute to the flat plateau in
$D(\omega)$ above
$\omega^*$.  
In other words, anomalous modes with characteristic length
smaller than
$l^*$ are not affected very much by the extra contacts, and the density of states is
unperturbed above a frequency $\omega^* \sim \delta z$.
This scaling is checked numerically in  Fig.~\ref{sans}. This prediction is in very good
agreement with the data up to $\delta z\approx 2$. 
 
\begin{figure}
\centering
\includegraphics[angle=270,width=7cm]{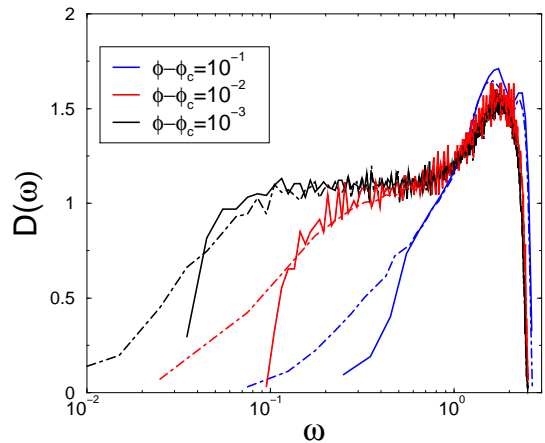}
\caption{Log-linear plot of the density of states for a 3-d system with N=1024 for three values of $\phi-\phi_c$ in the soft-sphere system (dotted line) and the system where the applied stress term has been removed (solid line). }
\label{compare}
\end{figure}

At frequencies lower than $\omega^*$ we expect the system to behave as a disordered, but not poorly-connected, elastic medium.  We expect that the vibrational modes will be similar to the plane waves of a continuous elastic body.  We refer to these modes as ``acoustic modes''.  Thus we expect $D(\omega)$ at small $\omega$ to vary as $\omega^{d-1}c^{-d}$, where $c(\delta z)$ is the sound speed at the given compression. This $c$ may be inferred from the bulk and shear moduli measured in the simulations \cite{durian,ohern,J}; one finds the transverse velocity $c_t \sim (\delta z)^{1/2}$, and the longitudinal velocity $c_l\sim \delta z^0$ in both three- and two- dimensions. Thus at low frequency $D(\omega)$ is dominated by the transverse acoustic modes and at $\omega=\omega^*$ the acoustic density of states is $\omega^{d-1}c_t^{-d}\sim\delta z^{d-1}\delta z^{-d/2}\sim \delta z^{d/2 -1}$.  For a three-dimensional system the acoustic density of states should be dramatically smaller than the plateau density of states.  Since there is no smooth connection between the two regimes we expect a sharp drop-off in $D(\omega)$ for $\omega<\omega^*$.  Such a drop-off is indeed observed, as seen for a three dimensional system in Fig.~(\ref{compare}). In fact, because of the finite size of the simulation,  no acoustic modes are apparent at $\omega<\omega^*$ near the transition. 

The behavior of such systems near the jamming threshold thus depend on the frequency $\omega$ at which they are observed. For $\omega>\omega^*$ the system behaves as an isostatic system, and for $\omega<\omega^*$ it behaves as a continuous elastic medium. Since the transverse and the longitudinal velocities do not scale in the same way, the presence of a unique cross-over in frequency leads to  the appearance of two distinct length scales $l_l$ and $l_t$, defined as $l_l\sim c_l \omega^*{}^{-1}$ and $l_t\sim c_t \omega^*{}^{-1}$. These lengths correspond  respectively to the wavelengths of the longitudinal and transverse acoustic modes at $\omega^*$. Note that since $c_l\sim \delta z^0$, one has  $l_l\sim l^*$. Interestingly, $l_t\sim \delta z^{-1/2}$ is the smallest system size for which acoustic modes exist.  For smaller system sizes, the lowest frequency mode is not a plane wave, but is an anomalous mode. $l_t$ can be observed numerically by considering the peak of the transverse structure factor at $\omega^*$ \cite{nagel}.

Our argument ignores the spatial fluctuations of $\delta z$.  If these fluctuations were spatially uncorrelated they would be Gaussian upon coarse-graining:  then the extra number of contacts $\Delta N_c$ in a subregion of size $L$ would have fluctuations of order  $L^{d/2}$. The scaling of the contact number that appears in our description is $\Delta N_c \sim L^{d-1}$ and is therefore larger than these Gaussian fluctuations for $d>2$. In other terms at the length scale $l^*$, where soft modes appear, the fluctuations in the number of contacts inside the bulk are negligible in comparison with the number of contacts at the surface.  Therefore the anomalous modes are not sensitive to fluctuations in coordination number in three dimensions near the transition.  In the discussion section we will argue that there are spatial anti-correlations in $z$, so that fluctuations also do not affect the extended soft modes in two dimensions. 

Note that these arguments do not preclude the existence of low-frequency localized modes that may appear in regions of small size $l<<l^*$, and that could be induced by very weak local coordination or specific arrangements of the particles.  The presence of such modes would increase the density of states at low-frequency. There is no evidence for their presence in the simulations of \cite{J}.

\section{ IV Effects of the applied stress on vibrational modes}

In this section we describe how the above simple description of $D(\omega)$ is affected by the presence of an applied stress. In general when a system of particles at equilibrium is formed, there are forces between interacting particles.  For harmonic soft spheres it leads to a non-vanishing first term in Eq.~(\ref{1}) $\frac{1}{4}\sum_{\langle  ij \rangle} (r_{ij}^{eq}-1) [(\delta\vec{R_j}-\delta\vec{R_i})^{\bot}]^2$, where we used $r_{ij}\approx 1$. This term is (a) negative for repulsive particles (b) proportional to the transverse relative displacement between particles in contact, (c) scales linearly with the pressure $p$ and therefore vanishes at the jamming transition.  The full dynamical matrix ${\cal D}$  can be written:

\be
{\cal D=M+M'}
\ee
where ${\cal M'}$ is written in tensorial notation in the footnote \footnote[5]{in three dimension we have ${\cal M'}_{ij}=-\frac{1-r_{ij}}{2r_{ij}} [ \delta_{\langle ij \rangle} (\vec{m_{ij}}\otimes \vec{m_{ij}}+ \vec{k_{ij}}\otimes \vec{k_{ij}}) + \delta_{i,j} \sum_{<l>}(\vec{m_{ij}}\otimes \vec{m_{ij}}+ \vec{k_{ij}}\otimes \vec{k_{ij}})$, where ($\vec{n_{ij}}$,$\vec{m_{ij}}$,$\vec{k_{ij}}$) is an orthonormal basis.}. The spectrum of $\cal D$ has {it a priori} no simple relation with the spectrum of $\cal M$. Because ${\cal M'}$  is much smaller than ${\cal M}$ near the transition, one can successfully use perturbation theory for the bulk part of the normal modes of $\cal{M}$. However perturbation theory fails at very low frequency, which is of most interest to us here.  In this region the spectrum of ${\cal M}$ contains the acoustic modes and the ``anomalous modes" forming the plateau. In what follows we estimate the change of frequency induced by the applied stress on these modes. We show that the relative correction to the plane-wave frequencies is very small, whereas the frequency of the anomalous modes can be changed considerably. Finally we show that these considerations lead to a correction to the Maxwell rigidity criterion.

\subsection{ Applied stress and acoustic modes}

Consider a plane wave of wave vector $k$.  As the directions $\vec n_{ij} $ are random, both the relative longitudinal and transverse displacements of this plane wave are of the same order: $[(\delta\vec{R_i}-\delta\vec{R_j})^\bot]^2\sim[(\delta\vec{R_i}-\delta\vec{R_j}).\vec n_{ij}]^2 \sim k^2$. Consequently the relative correction $\Delta E/E$ induced by the applied stress term is very small:

\be
\label{bb}
\frac{\Delta E}{E}\approx \frac{\frac{1}{2}\sum_{\langle  ij \rangle} (r_{ij}^{eq}-1)[(\delta\vec{R_j}-\delta\vec{R_i})^{\bot}]^2}{\sum_{\langle  ij \rangle} [(\delta\vec{R_j}-\delta\vec{R_i})\cdot \vec{n}_{ij}]^2}
\ee
since $r_{ij}^{eq}-1$ is proportional to the pressure $p$, while the others factors remain constant  as $p\rightarrow 0$, $\frac{\Delta E}{E}\sim p$, and is thus arbitrarily small near the jamming threshold \footnote[6]{In disordered systems the acoustic modes are not exact acoustic modes, see e.g. recent simulations in Lennard-Jones systems \cite{tanguy}. Therefore the correction due to the applied pressure on the frequency of the acoustic modes might be larger than what is expected from Eq.~(\ref{bb}). Nevertheless, we still expect the effect of applied pressure to be much smaller on the acoustic modes than on the anomalous modes.}.

\subsection{Applied stress and anomalous modes}
\label{ppp}
\begin{figure}

\centering
\includegraphics[width=5cm]{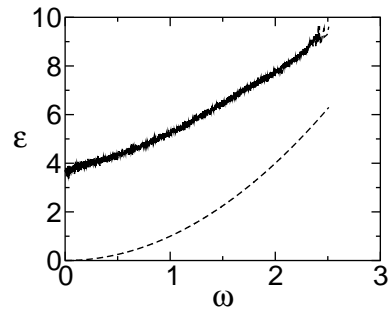}
\caption{Sum of the transverse terms (full curve) $\epsilon=  1/2 \sum_{\langle ij\rangle}[(\delta\vec{R_j}-\delta\vec{R_i})^{\bot}]^2$ and longitudinal terms (dotted curve) $\epsilon=  1/2\sum_{\langle ij\rangle}[(\delta\vec{R_j}-\delta\vec{R_i})\cdot \vec{n}_{ij}]^2$ for each mode of frequency $\omega$ at the jamming threshold in 3-dimensions. The longitudinal term is equal to the energy of the modes and vanish quadratically at 0 frequency. The transverse term converges toward a constant different from 0.}
\label{trans}
\end{figure}

For anomalous modes the situation is very different:  we expect the transverse relative displacements to be much larger than the longitudinal ones.  Indeed soft modes were built by imposing zero longitudinal terms, but there were no constraints on the transverse ones. These are the degrees of freedom that generate the large number of soft modes. The most simple assumption is that the relative transverse displacements are of the order of the displacements themselves, that is $\sum_{\langle ij\rangle}[(\delta\vec{R_j}-\delta\vec{R_i})^{\bot}]^2\sim \sum_i\delta\vec{R_i}^2=1$ for the anomalous modes that appear above $\omega^*$. This estimate can be checked numerically for an isostatic system where this sum is computed for all $\omega$. The sum of the transverse relative displacements converges to a constant when $\omega\rightarrow 0$ as assumed, see Fig.~(\ref{trans}). 

Finally we can estimate the scaling of the correction in the energy $\Delta E$ induced by the stress term on the anomalous modes: 

\be
\Delta E \sim - \sum_{\langle  ij \rangle} (1-r_{ij}^{eq}) [(\delta\vec{R_j}-\delta\vec{R_i})^{\bot}]^2 \sim -p
\ee
which is an {\it absolute} correction, and which can be  non-negligible in comparison with the energy E.

\subsection{ Onset of the anomalous modes}

We can now estimate the lowest frequency of the anomalous modes. The modes that appear at $\omega^*$ in the relaxed-spring system have an energy lowered by an amount on the order of $-p$ in comparison to the original system. Applying the variational theorem of the last section to the collection of slow  modes near $\omega^*$, one finds that there must be slow normal modes with a lower energy. That is,  the energy  $\omega_{AM}^2$ at which anomalous modes appear satisfies:
\be
\label{33}
\omega_{AM}{}^2\leq  \omega^*{}^2-A_2 p \equiv A_1 \delta z^2-A_2 p
\ee
where $A_1$ and $A_2$ are two positive constants. Thus coordination and stress determine the onset of the excess-modes. 

\subsection{Extended Maxwell criterion for the contact number under stress}

\begin{figure}
\centering
\includegraphics[width=7cm]{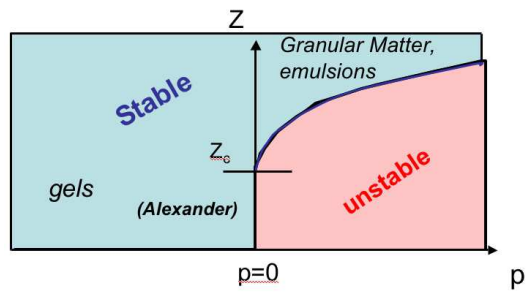}
\caption{Phase diagram of rigidity in terms of coordination number and pressure.  When $p>0$, the line separating the stable and unstable regions is defined by Eq.~(\ref{22}).  When the pressure is negative, any connected system can be rigid. }
\label{diag}
\end{figure}

From this estimate we can readily obtain a relation between the coordination and pressure that guarantees the stability of a system. There should be no negative frequencies in a stable system, therefore $\omega_{AM}>0$.  Thus in an harmonic system the right hand side of Eq.~(\ref{33}) must be positive:

\be
\label{22}
\delta z \geq C_0 p^{1/2}\equiv \delta z_{min}
\ee 
 where $C_0$ is a constant. This inequality, which must hold  for any spatial dimension, indicates how a system must be connected to counterbalance the destabilizing effect of the pressure.  A phase diagram of rigidity is represented in Fig.~(\ref{diag}). When $p=0$, the minimal coordination $z_c$ corresponds to the isostatic state: this is the Maxwell criterion. As we said, for spherical particles $z_c=2d$. When friction is present, one finds $z_c=d+1$. When $p>0$, Eq.~(\ref{22}) delimits the region of rigid systems: granular matter, emulsions lie above this line. As was shown by Alexander \cite{shlomon}, when $p<0$, even systems with many fewer contacts than required by the Maxwell criterion are rigid. These systems contain many soft modes as defined in Eq.~(\ref{3}), but they are all stabilized by the positive bracketed term of Eq.~(\ref{1}).  This is the case for example in a gel where polymers are stretched by the osmotic pressure of the solvent. Thus the network of reticulated polymers carries a negative pressure, which rigidifies the system and leads to a non-vanishing shear modulus. 

The relation (\ref{22}) is  verified in the simulations of \cite{J} where $\delta z\sim p^{1/2}$:  the numerical results are in agreement with an {\it equality} of Eq.~(\ref{22}).  Furthermore it appears in Fig.~(\ref{compare})  that $\omega^*>> \omega_{AM}$. In other words, the two opposite effects of pressure on the vibrations, that is (i) the increase in the coordination and (ii) the  addition of a negative term in the energy expansion, compensate \footnote[33]{ Assuming an exact compensation of these two terms lead to $\omega_{AM}=0$ in an infinite size system. In Fig.~(\ref{compare}) $\omega_{SM}$ is slightly different from zero. We believe that it is simply a finite size effect.}.  In the discussion section we justify why the system of \cite{J} is thus marginally stable even above the transition when $p>0$, and we furnish examples of dynamics that lead to such features.

\section{Discussion}

\subsection{Stronger constraints on $\delta z$} 

The simulations of \cite{J} show $\delta z \sim p^{1/2}$, thus potentially saturating the bound of Eq. (20), so that there are excess modes extending to frequencies much lower than $\omega^*$.  Here we furnish an example of dynamics that lead to such a situation.  Consider an initial condition where forces are balanced on every particle, but such that the inequality (\ref{22}) is not satisfied. Consequently, this system is not stable:  infinitesimal fluctuations make the system relax with the collapse of unstable modes. Such dynamics were described by Alexander in \cite{shlomon} as structural buckling events: these events are induced by a positive stress as occurs in the buckling of a rod; but here they take place in the bulk of an amorphous solid. These events {\it a priori} create both new contacts and decrease the pressure. When the bound of (\ref{22}) is reached, there are no more unstable modes. If the temperature is zero, the dynamics stop.  Consequently one obtains a system where Eq.~(\ref{22}) is an {\it equality}.  Therefore (i) this system is weakly connected (ii)  $\omega_{AM}\approx0$, so that there are anomalous modes very different from acoustic modes extending to zero frequency. 

In the simulations of Ref.~\cite{J} the relaxation  proceeds in the following way.  The system is initially in equilibrium at a high temperature.  Then it is quenched instantaneously to zero temperature.  At short time scales the dynamics that follows is dominated by the relaxation of the stable, high frequency modes. The main effect  is to restore approximately force balance on every particle.  At this point, if the inequality (\ref{22}) is satisfied, then the dynamics stop.  However, if it is not satisfied then we are in the situation of buckling as described above.  The pressure and co-ordination number continue to change until the last unstable mode has been stabilized.  At this point the bound of Eq. (20) is marginally satisfied, and there is no driving force for further relaxation.

The data of \cite{J} were obtained by gradually decreasing the
pressure from this initial state of zero temperature and nonzero
pressure.  The reduction of pressure causes some contacts to open.
The opening of these contacts tends to destabilize normal modes and
reduce their frequencies, while the reduction in pressure tends to
stabilize them.  If the particles simply spread apart affinely, the
destabilizing effect would be expected to dominate \footnote[33]{When
  the particle's radius decreases by an amount $\epsilon$, a certain
  fraction $e\sim g(1)\epsilon$ of contacts opens, where $g(r)$
  denotes the radial distribution function.  For harmonic particles we
  expect $g(1)\sim (\phi-\phi_c)^{-1}$, as we justify at the end of
  the paragraph. Thus using the fact that $p\sim (\phi-\phi_c)$ for harmonic
  particles \cite{J}, one obtains $e\sim\frac{\epsilon}{p}$.  On the
  other hand, the affine decompression lowers the pressure by an
  amount $\delta p\sim \delta \phi \sim \epsilon$. Thus the fraction
  $f$ of contacts that the system can afford to lose while staying
  stable follows, according to Eq.~(\ref{22}): $f= \frac{d(\delta
    z_{min})}{dp}\delta p\sim p^{-1/2} \delta p\sim
  \frac{\epsilon}{p^{1/2}}\ll e$. Therefore $f\ll e$ as claimed.  Thus
  if an affine reduction of packing fraction $\epsilon$ is imposed,
  far too many contacts open and the system is unstable. To conclude
  we justify $g(1)\sim (\phi-\phi_c)^{-1}$. This is related to
  well-known empirical facts of the force distribution: one has $P(F)
  dF\sim g(r) dr$. For harmonic particles $dr\sim dF$ and therefore
  $P(F)\sim g(r)$. When rescaled by $\langle F \rangle \sim p$, $P(F)$
  converges to a master curve with $P(F=0)\neq 0$ \cite{J,ohern}.
  This implies that $g(1)\sim p^{-1}\sim (\phi-\phi_c)^{-1}$
  \cite{ohern2}.}.  Thus we argue that some incremental buckling must occur as in
the initial temperature quench.  The buckling increases the contact
number and decreases the pressure until marginal stability is
achieved, so that the inequality of Eq.~(\ref{22}) is marginally
satisfied as the pressure decreases.

It is interesting to discuss further which systems, and which procedures, can have marginal stable states.  In repulsive short-range systems, we expect the situation of marginal stability that follows an infinite cooling rate to take place for a domain of the parameters of initial conditions $(\phi,T)$, located at high temperature and low density. This domain might stop at a finite $\phi$ even when the temperature is infinite. This is true for particles interacting with a gaussian potential, as was shown in simulations and  theoretical analysis on Euclidian random matrices \cite{parisi2}.  There most of the unstable modes vanish at a finite $\phi$ despite $T=\infty$. 

When the cooling rate is finite we expect that the relaxation does not stop when all the modes are stable. For example activated events can in principle lead to the collapse of anomalous modes.  These events {\it a priori} increase the connectivity and decrease the pressure further than the bound of Eq.~(\ref{22}), leading to $\omega_{AM} > 0$.  Interestingly hyper-quenched mineral glasses show a much larger number of excess modes \cite{angell} in comparison with normally cooled glasses, whereas annealed polymeric glasses show a diminished excess ofsuch modes \cite{duval}.

\subsection{Fluctuation of connectivity}

In Section III we argued that although fluctuations were negligible in three dimensions near the transition, they might have effects in two dimensions. Here we address these possible effects. In other systems such as ferromagnets the spatial fluctuations of eg. magnetic coupling lead to noticeable and sometimes striking effects \cite{Griffiths}.  It is natural to ask what analogous effects might be induced by fluctuations of contact number $z$ in the system of \cite{J} in two dimensions.  In a system with uncorrelated contacts in two dimensions, the modes in the weakly connected regions would be softer than those in the more strongly connected regions.  This would be expected to create modes with frequencies below the nominal $\omega^*$ in the system of relaxed springs of Section III, and would imply the presence of modes below the bound $\omega_{AM}$ in the original system. We argue that in the simulations of \cite{J} such fluctuations are negligible.  It is important to note that $z$ is not an uncorrelated random variable like the magnetic coupling mentioned above. For example, if $z$ were an 
uncorrelated random variable, one would expect that the fluctuations 
in the total number of contacts $N z$ would be of order $\sqrt{N}$. 
However, at the jamming threshold the system is isostatic and the 
number of contacts is precisely $2dN$; there are no fluctuations.  
More generally, the bound on  $\delta z$ in the last section applies not merely on average but rigorously to any subregion of size larger than $l^*$, since all sub-regions experience the same pressure $p$. Furthermore for a marginally stable system such as those of \cite{J}, the average $\delta z$ is given by the bound of Eq.~(\ref{22}). These two facts imply that the 
fluctuations of contact number are small. Therefore the conclusions of the last 
section remain valid, and no anomalous modes below the bound $\omega^*$ are 
expected in the relaxed-spring system due to fluctuations of $z$.

\subsection{Extension to non-harmonic contacts}

In the previous sections we considered harmonic interactions. Here we discuss the generalization of our argument to other potentials.  Ref.~\cite{J} explored several other interactions, notably the Hertzian interaction potential describing the compressive energy of two elastic spheres.  It corresponds to $\alpha = 5/2$ in Eq.~(\ref{opo}).  Ref.~\cite{J} observed a plateau in the density of states whose height  $D_0$ scales as $p^{-1/6}$.  They also observed a cutoff frequency $\omega^*$ varying as $p^{1/2}$.  In the Herztian case the quadratic energy of Eq. (7) becomes:
\be
\delta E= \frac{1}{2} \sum_{\langle  ij \rangle} (1-r_{ij})^{1/2}[(\delta\vec{R_j}-\delta\vec{R_i}).\vec n_{ij}]^2 
\ee
The new factor $(1-r_{ij})^{1/2}$ amounts to a spring constant 
$k_{ij}$ that depends on compression.  The contact force $f_{ij} = 
\partial \delta E/\partial r_{ij}$ evidently varies as 
$(1-r_{ij})^{3/2}$.  In what follows we neglect the fluctuations that exist between the contacts. This treatment is sufficient to recover the scaling results of \cite{J}. 

The new factor $(1-r_{ij})^{1/2}$ rescales the energy.  To account for this overall 
effect, we replace $(1-r_{ij})^{1/2}$ by its average 
$\langle (1-r_{ij})^{1/2}\rangle$.  Expressed in terms of contact 
forces, this factor is proportional to 
$\langle f_{ij}^{1/3}\rangle$.  Replacing $f_{ij}$ by its 
average, the factor becomes $\langle f_{ij}\rangle^{1/3}$.  This 
average is related to the pressure $p$, via $ p\approx 
\langle f_{ij}\rangle$.  Thus, in this approximation the overall effect is to rescale the 
energy by a factor $k(p) \sim p^{1/3}$. 
\be
\delta E= \frac{k(p)}{2} \sum_{\langle  ij \rangle} [(\delta\vec{R_j}-\delta\vec{R_i}).\vec n_{ij}]^2 
\ee

Apart from this prefactor, the energy and the dynamical matrix 
are identical to the harmonic case treated above.  Each normal 
mode frequency gains a factor $k^{1/2} \sim p^{1/6}$.  In the 
harmonic case the crossover frequency follows $\omega^* \sim \delta z$.  In 
the Hertzian case, it  gains the same factor $k^{1/2}$, so that $\omega^* 
\sim k^{1/2} \delta z$.  The bound on the lowest-frequency anomalous modes $\omega_{AM}$ still has the
form  
\be
\label{34}
\omega_{AM}{}^2 \leq  \omega^*{}^2 -A_2 p
\ee
For a marginally stable system we still have $\omega_{AM}=0$, which leads to an unaltered relationship between
$\omega^*$ and $p$: $\omega^* \sim p^{1/2}$.  Comparing with our previous estimate of $\omega^*$ we find $\delta z\sim p^{1/3}$. Furthermore, the plateau density of states $D_0$ has the dimension of an  
inverse frequency and thus gains a factor $p^{-1/6}$.  Since the 
harmonic $D_0$ had no dependence on $p$, the Hertzian $D_0(p)$ also 
should vary as $p^{-1/6}$.  The scaling behaviors seen in \cite{J} agree 
with these expectations.  These arguments may be
applied to general values of the interaction exponent $\alpha$.

Additional effects could in principle alter the low-frequency modes in the Hertzian case.  When harmonic springs are replaced by Hertzian ones, different contacts have different stiffness. This effect should be quantified in order to gain a more detailed understanding of the Hertzian case \cite{these}. 

\section{Conclusion}

In this paper we computed some vibrational properties of
weakly-connected amorphous systems. We introduced a frequency scale
$\omega^*$ above which such systems do not behave as continuous
elastic media, but as isostatic systems that are marginally stable. At
frequencies lower than $\omega^*$, the system acts as a continuous
solid, with low-frequency, acoustic, phonon modes. At frequencies
above $\omega^*$ the vibrational modes correspond to the anomalous
modes that constitute the vibrations of isostatic states.  As we
showed in Section \ref{to}, these anomalous modes are built from the
soft modes that appear in subsystems with free-boundary conditions.
The anomalous modes that appear at $\omega^*$ are characterized by a
length scale $l^*$. Interestingly $l^*$ does not appear directly in
any correlation function of the static structure, but rather in the
response fucntions of such systems. $l^*$ can be much larger than the
particle size and varies with the coordination number, as
$l^*\sim\omega^*{}^{-1}\sim \delta z^{-1}$.  Secondly we computed the
effect of applied stress on these anomalous modes. In a repulsive
system the stress has a strong effect that lowers the frequency of the
anomalous modes. Imposing that such modes are stable leads to a
generalization of the Maxwell criterion: the coordination must
increase non-analytically with pressure to compensate for the
destabilizing effect of compression. Finally, we discussed the
``structural buckling'' that occurs when anomalous modes collapse. We
use this concept to justify the marginal stability that follows
hyper-quenches in the simulations of repulsive, short-range systems of
\cite{J}.

The anomalous modes offer a new approach for understanding response and
transport in weakly-connected mechanical systems.  Knowledge of the
statistical properties of the anomalous and soft modes defined by
Eq.~(\ref{3}) is necessary to predict the acoustic and thermal
transport of e.g. the simulated system of \cite{J}.  These modes also
provide an alternative view of relaxation, via buckling of compressed
anomalous modes.  This buckling picture provides an intriguing
contrast to the local cage-escape picture commonly used to describe
these relaxations.  This approach may be more broadly applicable to
glasses. Such glasses exhibit a large excess of low frequency modes
like our marginally jammed system; this suggests that they behave like
weakly-connected mechanical systems at short length scales.  This
resemblance raises the hope that our approach may explain anomalous
phenomena in the transport, response, relaxation and aging of
structural glasses.

 We  thank J.P Bouchaud, J-L Barrat, Bulbul Chakraborty, L.E. Chayes, P.G. De Gennes, X. Leonforte, Andrea Liu, Corey O'Hern, Jennifer Schwartz, and A. Tanguy for helpful discussions. We also acknowledge the support of the CFR fellowship, MRSEC DMR-0213745 and DOE grant DE-FG02-03ER46088.
 
\section{Appendix: Spatial  distribution of the soft modes}
\label{geo}

In our argument we have assumed that  when $q\sim L^{d-1}$ contacts were cut along a hyperplane in an isostatic system, there was  a vector space of dimension $q'=aq$ which contains only extended modes, and that $a$ does not vanish when $L\rightarrow \infty$.  A normalized mode $ |{\bf \delta R}\rangle$ was said to be extended if $\sum_i\sin^2(\frac{x_i \pi}{L}) \langle i|{\bf \delta R }\rangle^2 > e_{0}$, where $e_{0}$ is a constant, and does not depend on L. Here we show how to chose  $e_0>0$ so that there is a non-vanishing fraction of extended soft modes. We build the vector space of extended soft modes and  furnish a bound to its dimension.  

Let us consider the linear mapping $\cal G$ which assigns to a displacement field $|{\bf \delta R} \rangle$ the displacement field $ \langle i| {\cal G} {\bf\delta R }\rangle= \sin^2(x_i\pi/L) \delta \vec {R }_i$.  For any soft mode $|{\bf \delta R}_\beta \rangle$ one can consider the positive number $a_\beta\equiv {\langle \bf \delta R_\beta}|{\cal G}|{\bf\delta R}_\beta \rangle \equiv \sum_i   \sin^2(x_i\pi/L) \delta \vec {R }_{i,\beta}^2$. We build the vector space of extended modes by recurrence:  at each step we compute the $a_\beta$ for the normalized soft modes, and we eliminate the soft mode with the minimum $a_\beta$. We then  repeat this procedure in the vector space orthogonal to the soft modes eliminated. We stop the procedure when $a_\beta>e_{0}$ for all the soft modes $\beta$ left.  Then all the modes left are extended according to our definition. We just have to show that one can choose $e_{0}>0$ such that when this procedure stops, there are $q'$ modes left, with $q'>aq$, with $a>0$.  In order to show that, we introduce the following overlap function:

\be
\label{lei}
f(x) dx \equiv  q^{-1} \sum_{\beta=1,...,q}\sum_{x_i\in [x, x+ dx]} [\delta \vec{R}_{i,\beta}]^2
\ee

The sum is taken on an orthonormal basis of soft modes $\beta$ and on all the particles whose position has a coordinate $x_i\in [x, x+ dx]$. $f(x)$ is the trace of a projection, and is therefore independent of the orthonormal basis considered. $f(x)$  describes the spatial distribution of the amplitude of the soft modes. The $|\delta {\bf R}_\beta\rangle$ are normalised and therefore:

\be
\int_0^L f(x)dx=1
\ee

\begin{figure}

\centering
\includegraphics[width=7cm]{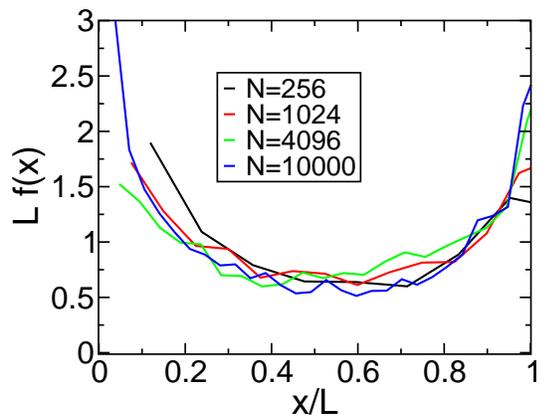}
\caption{The overlap function $f(x)$ as defined in Eq.~(\ref{lei}) for different system sizes in three dimensions. Soft modes were created from isostatic configurations as described in Fig.~\ref{softfig}}
\label{f}
\end{figure} 

We have examined soft modes made from  configurations at the jamming transition found numerically in \cite{J}.  The overlap function $f(x)$ was then computed for different system sizes $L$. These are shown in Fig.~\ref{f}. It appears from Fig.~(\ref{f}) that i) when $f(x)$  is rescaled  with the system size it collapses to a unique curve, and ii) this curve is bounded below by a constant $c$ ($c \approx 0.6$).  Consequently one can bound the trace of $\cal G$: $tr  {\cal G} =\int_0^L q f(x) \sin^2(x)> q c/2$. On the other hand one has $tr {\cal G} =\sum_{\beta = 1}^q a_\beta$, where the sum is made on the orthonormal basis we just built in the previous paragraph. This sum can be divided into contributions from the non-extended states and extended states:
\be
\sum_{\beta = 1}^q a_\beta =
\sum_{\beta = 1}^{q-q'} a_\beta + 
\sum_{\beta = q-q'+1}^q a_\beta.
\ee
By construction all the $a_\beta$ in the first
sum are smaller than $e_0$, while all the $a_\beta$ in the (first or) second are
smaller than 1.  Thus 
$\sum_{\beta = 1}^q a_\beta < (q-q')e_0 + q'$.  Combining, we have
$(q-q')e_0 + q' > qc/2$, or

\be
q' > q (c/2 - e_0) / (1 - e_0)
\ee

Evidently for all $L$, $q'$ remains a nonvanishing fraction of $q$ for any fixed
threshold
$e_0$  such that $e_0<c/2$, as claimed.

\end{document}